\documentstyle[12pt]{article}
\begin{document}
\newcommand{\beq}{\begin{equation}}\newcommand{\oper}[4]{\mbox{$O_{#1}^{#2}(#3,#4)$}}
\newcommand{\eeq}{\end{equation}}
\newcommand{\intif}{\int_{-\infty}^{\infty}}
\newcommand{\sid}{\mbox{$\psi^{\dagger}$}}
\newcommand{\sib}{\mbox{$\overline{\psi}$}}
\newcommand{\il}{\int_{-\Lambda}^{\Lambda}}
\newcommand{\ie}{\int_{0}^{\Lambda}}
\newcommand{\iT}{\int_{0}^{2\pi}d\theta}
\newcommand{\iK}{\int_{\Lambda /K_F}^{\pi -\Lambda /K_F}}
\newcommand{\si}[2]{\mbox{$\psi_{#1}(#2)$}}
\begin{center}
{\large Effective Field Theory in Condensed Matter Physics}\\
R.Shankar \\
Sloane Physics Lab\\
Yale University \\
New Haven CT 06520\\
\end{center}
\footnote{Based on the lecture given at the Boston Colloqium for the Philosophy of Science, March 1-3, 1996.}
I am presumably here to give you my perspective on quantum field theory from the point of view of a condensed matter theorist. I must begin with a disclaimer, a warning that I may not be really representing anyone but myself, since I find myself today working in condensed matter after a fairly torturous route. I must begin with a few words on this, not only since it will allow you to decide who else , if any, I represent, and also because my past run ins with field theory will parallel that of many others from my generation. 

I started life as an electrical engineer, as Professor Schweber said in the introduction. First I want to thank him for referring to me as a young theorist. As an electrical engineer,  the only field theorists  I  knew about  as an undergraduate were Bethe and Schwinger, who had done some basic work on wave guides. When I graduated, I  switched to physics and was soon working with Geoff Chew at Berkeley on particle physics. In those days, (sixties and early seventies), the community was split into two camps: the field theorist and the S-matricists and Geoff was the high priest of the latter camp.  The split was therefore not between field theory and effective field theory, but field theory and no field theory. The split arose because, unlike in QED, where everyone agreed that the electron and the proton go into the Lagrangian, get coupled to the photon, and out comes the hydrogen atom as a composite object, with strong interactions the situation seemed more murky.   Given any three particles, you could easily view one as the bound state of the other two because of the large binding energies. The same particles also played the role of mediating the forces. So the problem was not just, "What is the lagrangian?", but "Who goes into it and who comes out?". So Geoff and his camp decided to simply make no reference to Lagrangians and deal with just the S-matrix. Now Geoff was a very original thinker and deeply believed in this view. For example, he would refer to the pomeron, a pole in the complex angular momentum plane, as The Pomeron, in very concrete terms, but refer to the photon as the Photon Concept, even in a brightly lit room room bursting with photons.  I must  emphasize  that Geoff bore animosity towards field theory or  field theorists-- he greatly admired Mandelstam who was a top notch field theorist- but he would not tolerate any one confusing S-matrix ideas with field theory ideas. As you will hear from Professor Kaiser, this was easy to do.  So I had a choice: either struggle and learn field theory and run the risk of blurting out some four letter word like $\phi^4$ in Geoff's presence or simply eliminate all risk by avoiding the subject  altogether. Being a great believer in the principle of least action, I chose the latter route. Like Major Major's father in Catch 22, I woke up at the crack of noon and spent eight and even twelve hours a day not learning field theory and soon I had not learnt more field theory than anyone else in Geoff's group and was quickly moving to the top. 

Soon it was time to graduate.  By this time great changes had taken place. Clear evidence for quarks appeared at SLAC and with discovery of asymptotic freedom, QCD emerged as a  real possibility for strong interactions. Geoff then packed me off to Harvard, to the Society of fellows. I am sure he knew I would be mingling with the field theorists, but it did not bother him, which is why students like me admire him so much. I view his sending me off to Harvard   as  akin to Superman's father putting him in a capsule and firing him towards earth. However, unlike Superman who landed in a planet where he could beat the pants off everyone around him, I landed at Harvard, where  everyone knew more field theory than I did. I remember  one time when I was learning QED, the chap who came to fix the radiator, also stopped to help me fix my Coulomb gauge. Anyway I had very good teachers, and   many of my mentors from that time are at this conference. Pretty soon I went to the Coop, bought a rug for my office and was sweeping infinities under it- first a few logarithms and pretty soon, quadratic ones. So that is how I spent my winters in Cambridge: with a nice warm rug with with plenty of Logs under it. 

My stay was nearly over, when one day Ed Witten said too me, " I just learnt  a new way to find exact S-matrices in two dimensions invented by Zamolodchikov and I want to extend the ideas to supersymmetric models. You are the S-matrix expert aren't you? Why don't we work together?" I was delighted. All my years of training in Berkeley gave a tremendous advantage over Ed -- for an entire week. Anyway, around that time I moved to Yale and 
 learnt from Polyakov, who was visiting the US, that according to Zamolodchikov, these S-matrices were also the partition functions for Baxter-like  models. I could not resist plunging into statistical mechanics and never looked back.  Over the years I moved over to many-body physics, where I find myself now. 

Despite all this moving, I never stopped admiring or using field theory and have a unique perspective that can only be attained by achieving mediocrity in an impressive number of areas. To me field theory does not necessarily mean a Lorentz invariant, renormalizable theory in the continuum,  it means any theory that takes the form of a functional integral over an infinite number of degrees of freedom. 

I will now proceed to give my views on effective field theories, by which one means a field theory which works only down to some length scale or up to some energy scale.  Often people using effective field theories feel they are somehow compromising. I will now make my main point {\em Even when one knows the theory at a microscopic level, there is often a good reason to deliberately move away to an effective theory.} 
Now everyone  knows the microscopic Lagrangian in condensed matter: you just need electrons, protons and the coulomb interaction.  Everything we know follows from this. There is no point getting any more fundamental, it makes no difference that the photon is part of a bigger gauge group, or the electron and its neutrino form a doublet which itself is part of a family: we have no family values.  As long as you start with the right particle masses and charges, nothing matters. On the other hand,  progress is often made by moving away from the fundamental and looking for effective theories. 
I will illustrate my view with one example because  I am most familiar with since I have been working on it for some years and because some particle physicists like Polchinski and Weinberg have been interested in some aspects of this problem. Towards the end I will mention a few more.

The technique I am going to use is the Renormalization Group (RG). You will hear about in greater detail from Michael Fisher and others. I am going to give you a caricature. Imagine that you have some problem in the form of a partition function

\beq
Z(a,b) = \int dx \int dy e^{-a(x^2 + y^2)} e^{-b (x+y)^4}
\eeq
where $a,b$ are the parameters. Suppose now that you are just interested in $x$, say in its fluctuations.  Then you have the option of integrating out $y$ as follows

\begin{eqnarray}
Z(a', b' ...) &= &\int dx \left[ \int dy e^{-a(x^2 + y^2)} e^{-b (x+y)^4} \right] \nonumber \\ 
& \equiv       &     \int dx e^{-a'(x^2 )} e^{- b'x^4 - c' x^6 + ...}
\end{eqnarray}
 where $a'$, $b'$ etc., define the parameters of the effective field theory  for $x$.   These parameters will reproduce exactly the same averages for $x$ as the original  ones. This evolution of parameters with the elimination of uninteresting degrees of freedom, is what we mean these days by renormalization, and as such has nothing to do with infinities; you just saw it happen in a problem with just two variables. 

Notice that to get the effective theory we need to do a nongaussian integral. This can only be done perturbatively. At the simplest "tree Level", we simply drop $y$ and find $b'=b$. At higher orders, we bring down the nonquadratic exponential and integrate term by term and generate effective interactions for $x$. This procedure can be represented by Feynman graphs with the only difference that variables in the loop are limited  to the ones being eliminated.

Why do we do this? We do this because certain tendencies of $x$ are not so apparent when $y$ is around, but surface to the top, as we zero in on $x$. For example, a numerically small term can grow in size as variables as we eliminate the unwanted variables and guide us towards the correct low energy structure. This will now be demonstrated with an example. 

Consider a system of nonrelativistic spinless fermions in two space dimensions. The one particle hamiltonian is 
\beq
H = {K^2 \over 2m} - \mu
\eeq
where the chemical potential $\mu$  is introduced to make sure we have a finite density of particles in the ground state: all levels  up the Fermi surface, a circle defined by

\vspace*{-.5in}

\let\picnaturalsize=N
\def\picsize{3.0in}
\def\picfilename{Fig1.eps}
\ifx\nopictures Y\else{\ifx\epsfloaded Y\else\input epsf \fi
\let\epsfloaded=Y
\centerline{\ifx\picnaturalsize N\epsfxsize \picsize\fi \epsfbox{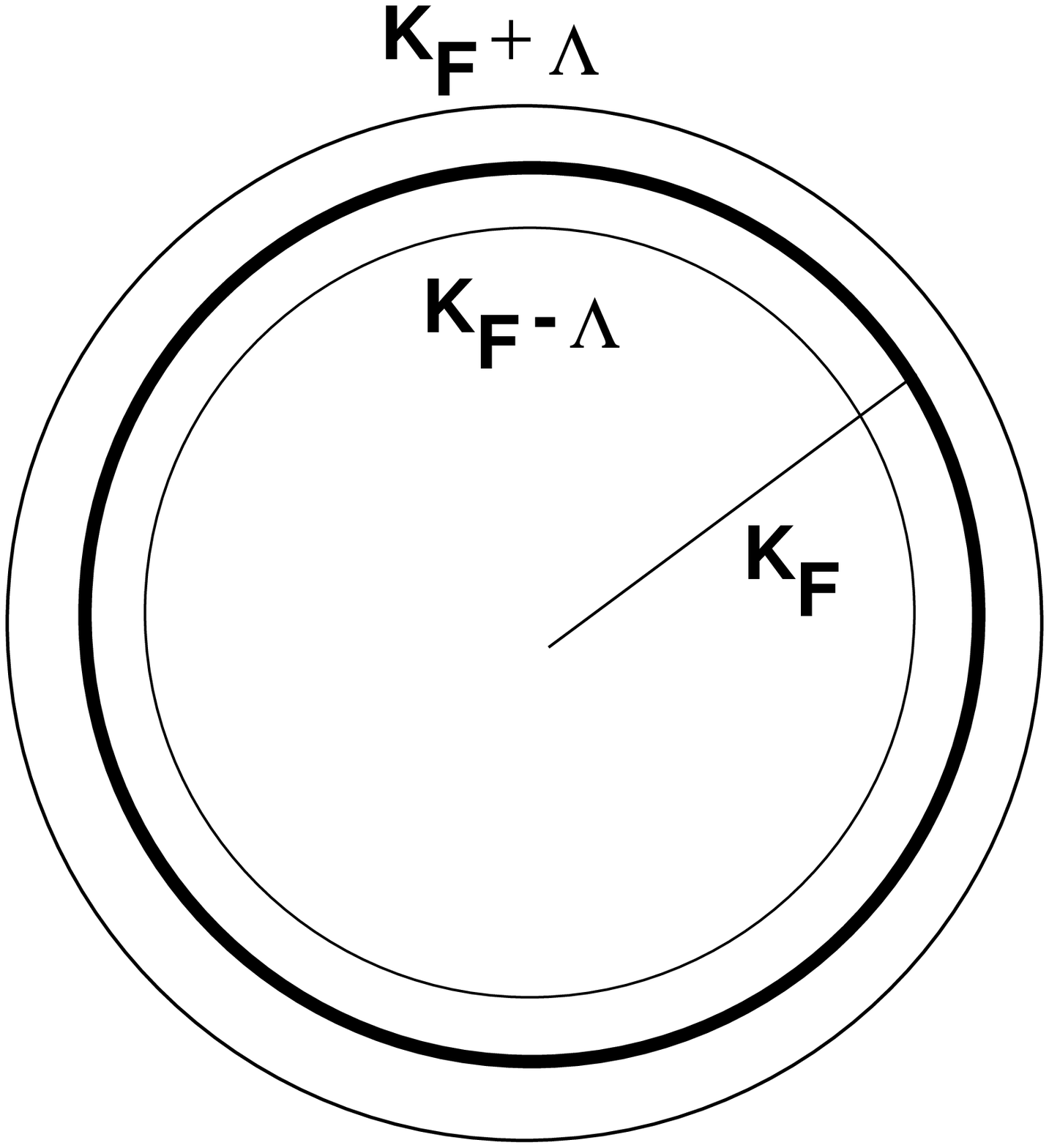}}}\fi

\vspace*{-.5in}

{\em {\bf Fig 1.}  The Fermi surface in two dimensions. The low energy modes lie within an annulus of thickness $\Lambda$ on either side of it.}\\
\newpage
\beq
K^{2}_{F} /2m = \mu 
\eeq 
 are now occupied and form a circle. 

Notice that this system has gapless excitations. You can take an electron just below the Fermi surface and move it just above, and this costs as little energy as you please. So one question one can ask is if this will be true when interactions are turned on. 

So we are going to answer this using the RG. Let us first learn how to do it for noninteracting fermions.  To understand the low energy physics, we take a band of of width $\Lambda$  on either side of the   Fermi surface. This is the first great difference between this problem and the usual ones in relativistic field theory and statistical mechanics. Whereas in the latter examples low energy means small momentum, here it means small deviations from the Fermi surface. Whereas in these  older problems we zero in on the origin in momentum space, here we zero in on a surface.

Let us cast the problem  in the form of a euclidean path integral: 
\beq
Z= \int  d\psi d\overline{\psi} e^{S_0}\\
\eeq
where 
\beq
S_0= \int_{0}^{ 2\pi}d\theta 
\intif d\omega \il dk \overline{\psi} (i\omega -
vk)\psi
\label{Z1}
\eeq
and $v$ is the Fermi velocity, which gives the derivative of the energy with radial momentum on the Fermi surface and $k = $K$ - K_F$.

Let us now perform mode elimination and reduce the cut-off by a factor $s$. Since this is a gaussian integral, mode elimination just leads to a multiplicative constant we are not interested in. So the result is just the same action as above, but with $|k| \le \Lambda$. Let us now do make the following additional transformations:

\begin{eqnarray}
(\omega ', k')                          &=& s(\omega , k)
\nonumber \\
(\psi ' , \overline{\psi}' )            &=& s^{-3/2} (\psi
 , \overline{\psi} ).
\label{rescale}
\end{eqnarray}

When we do this, the action and the phase space all return to their old values. So what? Recall that our plan is to evaluate the role of  quartic interactions in low energy physics as we do mode elimination.    Now what really matters is not the absolute size of the quartic term, but its size relative to the quadratic term.  By keeping fixed the size of the quadratic term and the phase space, as we do mode elimination,  this evaluation becomes very direct: if the quartic coupling grows, it is relevant;  if it  decreases, it is irrelevant, and it it stays the same it is marginal.

Let us now turn on a generic four-Fermi interaction: 
\beq
S_4 =   \int \overline{\psi}(4) \overline{\psi}(3) \psi
(2)\psi(1) u(4,3,2,1)\label{s4}
\eeq
where $\int $ is a shorthand:
\beq
\int \equiv \prod_{i=1}^{3} \int{d \theta_{i}}\int_{-
\Lambda}^{\Lambda} dk_{i} \intif d\omega_{i}
\eeq

At the tree level, we simply keep the modes within the new cut-off, rescale fields, frequencies  and momenta , and read off the new coupling. We find 

\beq
u'(k,\omega , \theta ) = u(k/s, \omega /s, \theta)
\label{tree}
\eeq

This is the evolution of the coupling function. To deal with constants, we expand the function in a Taylor series (schematic) 

\beq
u = u_o + k  u_1  + k^2  u_2 ...
\end{equation}
where $k$ stands for all the $k$'s  and $\omega$'s. An expansion of this kind is possible since 
 couplings in the Lagrangian are nonsingular in a problem with short range interactions.  

If we now make such an expansion and compare coefficients, we find that  $u_0$ is marginal and the rest are irrelevant, as is any coupling of more than four fields. 
Now this is exactly what happens in  $\phi^{4}_{4}$. The difference here is that we still have dependence on the angles on the Fermi surface: 
$$u_0 = u(\theta_1 , \theta_2 , \theta _3 , \theta_4 ) $$

Therefore in this theory you are going to get coupling functions and not a few coupling constants. 

Let us analyze this function. Momentum conservation should allow us to eliminate one angle. Actually it allows us more because of the fact that these momenta do not come form the entire plane, but  very thin annulus near $K_F$. 
Look at the left half of Figure 2. Assuming that the cutoff has been reduced to the thickness of the circle in the figure, it is clear that if two points  $1$ and $2$ are chosen from it to represent the incoming lines in a four point coupling, the outgoing ones are forced to be equal to them (not in their sum, but individually) up to a permutation, which is irrelevant for spinless fermions. Thus we have in the end just one function of the angular difference: 

\vspace*{-.5in}

\let\picnaturalsize=N
\def\picsize{3.0in}
\def\picfilename{Fig2.eps}
\ifx\nopictures Y\else{\ifx\epsfloaded Y\else\input epsf \fi
\let\epsfloaded=Y
\centerline{\ifx\picnaturalsize N\epsfxsize \picsize\fi \epsfbox{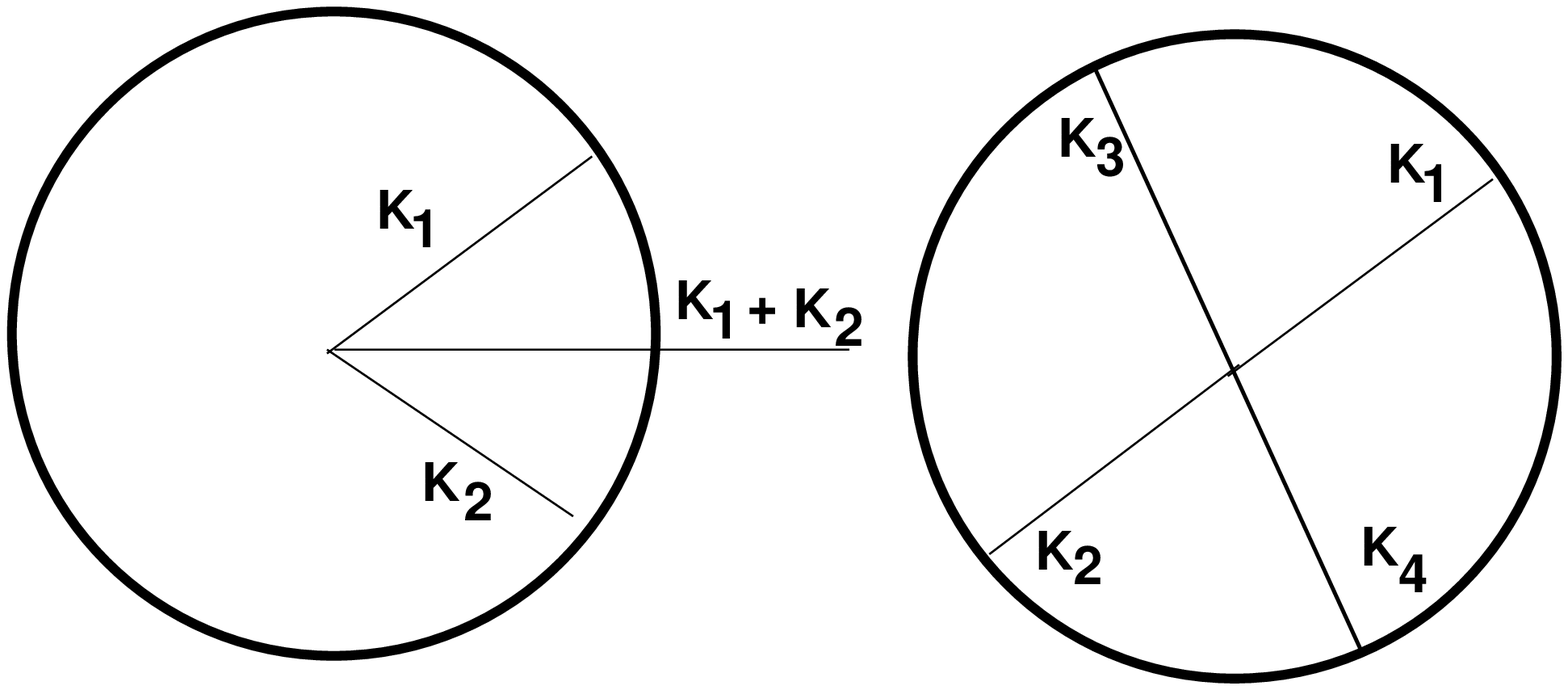}}}\fi
\vspace*{-.5in}

{\em {\bf Fig. 2} Kinematical reasons why momenta are either conserved pairwise or restricted to the BCS channel.}\\

\beq
u(\theta_1 , \theta_2 , \theta _1 , \theta_2 ) = F(\theta_1 - \theta_2 )  \equiv F(\theta ).
\eeq
About forty years ago Landau came to the very same conclusion that a Fermi system at low energies would be described by one function defined on the Fermi surface. He did this without the benefit of the RG and for that reason, some of the leaps were hard to understand.  Later detailed diagrammatic calculations justified this picture. The RG  provides yet another way to understand it. It also tells us other things as we will now see.  

The first thing is that the final angles  are not slaved to the initial ones if the former are exactly  opposite, as in the right half of Figure 2. In this case, the final ones can be anything, as long as they are opposite to each other. This leads to one more set of marginal couplings in the BCS channel, called

\beq
u(\theta_1 , \theta_1 , \theta _3 , -\theta_3 ) = V(\theta_3 - \theta_1) \equiv  V(\theta ).
\eeq

The next point is that $F$ and $V$ are marginal at tree level, we have to go to one loop to see if they are still so. So we draw the usual diagrams shown in Figure 3. We eliminate an  infinitesimal momentum slice of thickness $d\Lambda$ at $k= \pm \Lambda$.

\vspace*{-.25in}

\let\picnaturalsize=N
\def\picsize{4.0in}
\def\picfilename{Fig3.eps}
\ifx\nopictures Y\else{\ifx\epsfloaded Y\else\input epsf \fi
\let\epsfloaded=Y
\centerline{\ifx\picnaturalsize N\epsfxsize \picsize\fi \epsfbox{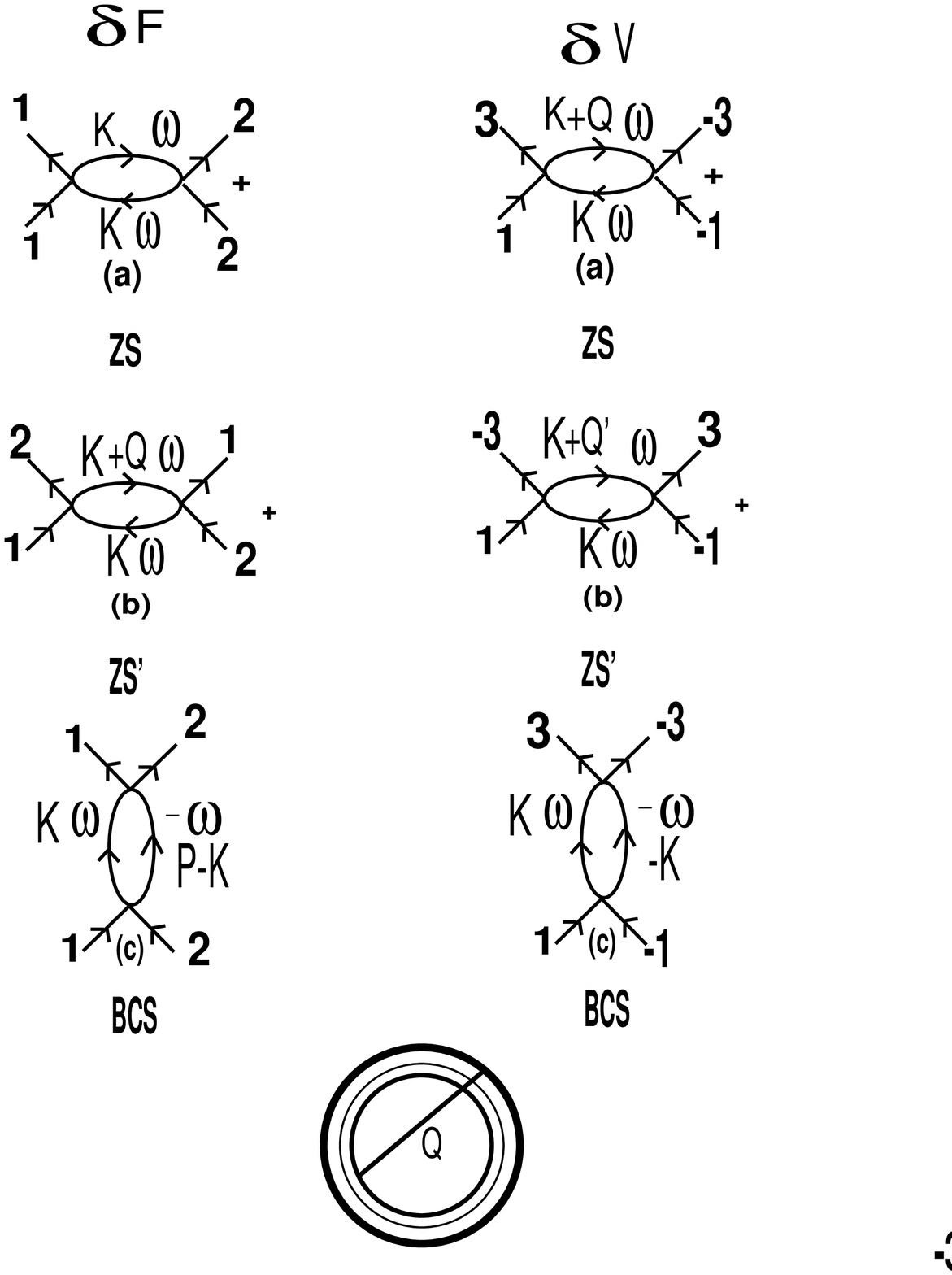}}}\fi
{\em {\bf Figure 3} One loop diagrams for the flow of $F$ and $V$.}\\

\newpage
These diagrams are like the ones in any quartic field theory, but each one behaves differently from the others and its its traditional counterparts. 
Consider the first one (called ZS) for $F$. The external momenta have zero frequencies and lie of the Fermi surface since $\omega$ and $k$ are irrelevant. The momentum transfer is exactly zero. So the integrand has the following schematic form:

\beq
\delta F 
\ \simeq
\int d{\theta} \int dk d\omega \left(
\frac{1}{(i\omega - \varepsilon(K))}\ \frac{1}{(i\omega  -
 \varepsilon(K'))} \right)
 \label{denom}
 \eeq
The loop momentum $K$ lies in one of the two shells being eliminated.  Since there is no energy difference between the two propagators, the poles in $\omega$ lie in the same half-plane and we get zero, upon closing the contour in the other half-plane. In other words, this diagram can contribute if it is a particle-hole diagram, but given zero momentum transfer we cannot  convert a hole at $-\Lambda$ to a particle at $+\Lambda$. In the ZS' diagram, we have a large momentum transfer, called $Q$ in the inset at the bottom. This is of order $K_F$ and much bigger than the radial cut-off, a phenomenon unheard of in say $\phi^4$ theory, where all momenta and transfers are of order $\Lambda$. This in turn means that the loop momentum is not only restricted in the direction to a sliver $d\Lambda$, but also in the angular direction in order to be able to absorb this huge momentum and land up in the other shell being eliminated. So we have $du \simeq dt^2$, where $dt = d\Lambda /\Lambda$. The same goes for the BCS diagram. 

Let us now turn to the renormalization of $V$. The first two diagrams are useless for the same reasons as before, but the last one is special. Since the total incoming momentum is zero, the loop  momenta are equal and opposite and no matter wht direction $K$ has, $-K$ is guaranteed to lie in the same shell. However the loop frequencies are now equal and opposite so that the poles now lie in opposite directions. We now get a flow (dropping constants)

\beq
{du(\theta_{1} - \theta_{3} ) \over dt} = - \cdot \int d\theta
u(\theta_{1}- \theta )\
u(\theta - \theta_{3} )
\label{bcs}
\eeq

Here is an example of a flow equation for a coupling function. However by expanding in terms of angular momentum eigenfunctions we get an  infinite number of flow equations

\beq
\frac{du_m}{dt} = - u_{m}^2.\label{bcsrg}
\eeq

one, for each coefficient. What these equations tell us is that if the potential in angular momentum channel $m$ is repulsive, it will get renormalized down to zero ( a result derived many years ago by Anderson and Morel) while if it is attractive, it will run off, causing the BCS instability.  This is the reason the $V$'s are not a part of Landau theory, which assumes we have no phase transitions.

Not only did Landau say we could describe Fermi liquids with an $F$, he also managed to compute the soft response functions of the system in terms of the $F$ function even when it was large, say $10$ in dimensionless units. The RG also gives us one way to understand this. As we introduce a $\Lambda$ and reduce it, a small parameter enters, namely $\Lambda /K_F$. This plays the role of $1/N$, $N$ being the number of species and the same diagrams that dominate  the $1/N$ expansion also dominate  here.  Recall that the action of the noninteracting fermions had the form of an integral over one-dimensional fermions, one for each direction. How many fermions are there? It is not infinity since each direction has zero measure. It urns out that in any calculation, at any given $\Lambda$, the number of fermions is $K_F/\Lambda$. If you imagine dividing the annulus into patches of size $\Lambda$ in the angular direction also, each patch carries an index and contributes to one species.  Landau theory is just the $N=\infty $ limit. The effective field theory has additional  symmetries that allows for its solution. 

A long  paper of mine  explains all this,  as well as how it is to be generalized to anisotropic Fermi surfaces and Fermi surfaces with additional special features and consequently additional instabilities.   

The general scheme was proposed many years ago by Anderson. 

Polchinski independently analyzed the isotropic Fermi liquid ( not in the same detail, since it was a just paradigm for him) and Weinberg independently derived the flow equations for the anisotropic superconductor.

I will now quickly discuss  a few other examples of effective field theories in condensed matter.  Consider spinless fermions on a  one dimensional lattice described by a hamiltonian: 
\begin{equation} 
H = - {1 \over 2} \sum_n \psi^{\dag} (n+1) \psi (n) + H.C.\label{micro}
\end{equation}

Upon Fourier transforming, we get the dispersion relation
\begin{equation} 
E(K) = - \cos K \hspace {2in} -\pi \le K \le \pi  .
\end{equation}
depicted in Figure 4.

\vspace*{-1in}

\let\picnaturalsize=N
\def\picsize{3.0in}
\def\picfilename{fig4.eps}
\ifx\nopictures Y\else{\ifx\epsfloaded Y\else\input epsf \fi
\let\epsfloaded=Y
\centerline{\ifx\picnaturalsize N\epsfxsize \picsize\fi \epsfbox{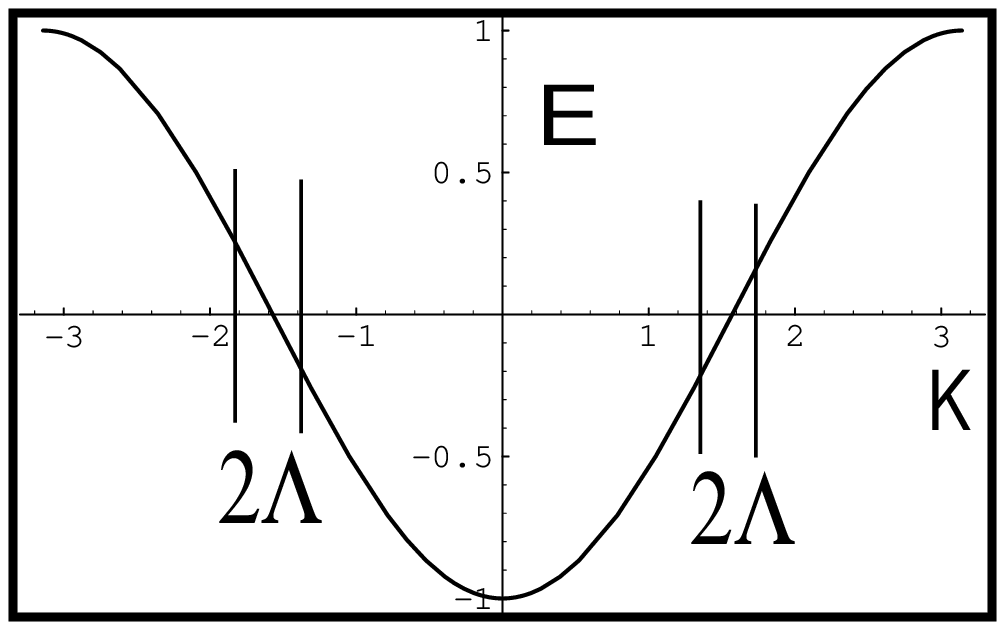}}}\fi

\vspace*{-1in}
{\em {\bf Figure 4} The energy momentum relation for free fermions on a linear lattice.}\\

 If we now linearize this near the Fermi points at $\pm K_F$ and introduce a  $k$ which measures  $K-K_F$
(and not$ |K|-K_F$ )  we find

$$ H_{eff} = \int_{-\Lambda}^{\Lambda}dk \left[  k\psi^{\dag}_R (k) \psi_R (k) - k\psi^{\dag}_L (k) \psi_L (k)  \right]
$$
which we recognize to be the beautiful one dimensional Dirac hamiltonian with $E = \pm k$. Thus from the original microscopic hamiltonian Eqn.(\ref{micro}), which even its mother would not call beautiful, we have obtained a beautiful effective theory. We can add many interactions, solve them by bosonization {\em a la} 
Luther and Peschel or Coleman.

Consider next the problem studied by Haldane, a one-dimensional spin antiferromagnetic Heisenberg chain with  

$$ 
H = \sum_n \vec{S} (n) \cdot \vec{S} (n+1) 
$$

This already an effective interaction , which can be traced back to the  coulomb interaction.  
Haldane wanted to show that contrary to naive expectations, this theory had two distinct limits as $S\to \infty$, depending on whether $S$ was integer or half-integer. He showed  that if we focus on the low energy modes of this model, we get the partition function for the nonlinear sigma model with a theta parameter equal to $2\pi S$:

$$ Z= \int \left[ dn \right] \exp \left[ - {1 /over 2g} (\nabla n )^2 + 2\pi iS W\right]
$$
where $W$ is the winding number, which is necessarily integer. It follows that if $S$ is an integer, the topological term makes no difference.  On the other hand, for half-integer values, it produces an alternating sign depending upon the instanton number. 

Sometime later I considered the effect of adding holes to this model. Now a hole is like a spinless fermion since you can only have zero or one hole per site, and the spinless fermion is equal to a Dirac field. So we get a sigma model coupled to a Dirac fermion. Just as the sigma model mimics QCD, this one mimics QCD coupled to massless quarks. In particular the massless fermions make the value of the topological coefficient irrelevant, which translates into the fact that these holes make all large $S$ systems behave in one way, as compared to two.

Lastly, let us recall that the Ising model near criticality can be represented by a Majorana field theory with action:

\beq
S = \int \overline{\psi} (i\ \partial +m )\psi  d^2 x
\eeq
where $m$ measures the deviation from criticality. If we now consider  a random bond Ising model in which the bond vary around the critical value, we an Ising model with a position dependent mass term. To find the quenched averages in this model, one uses the replica trick and obtains the $N=0$ component Gross-Neveu model, as shown by Dotsenko and Dotsenko. The amazing thing is that this model was 
originally  invented for $N\to \infty$. and we see it being useful down to $N=0!$.

To conclude, my main message is that very beautiful quantum field theories arise in condensed matter physics as effective theories and in addition to their beauty, effective field theories are also very effective in answering certain questions that the more microscopic versions cannot. 
\begin{center}
{\em {\bf Bibliography}}
\end{center}
Due to the informal nature of this talk, references will also be given in a similar vein and will not be exhaustive and limited to those that were referred to in this talk. Given the long history of the subject a complete listing of papers is not possible here.  For details on the RG applied to fermions see R.Shankar, Rev. Mod. Phys., Vol 66, 129, 1994. 
See also J. Polchinski , Proceedings of the 1992 TASI Elementary Particle Physics, Editors J. Polchinski and J.Harvey, World Scientific, 1992, and  S. Weinberg, Nucl. Phys., B413, 567, 1994. There has been a lot of subsequent literature on applications of these ideas to non-Fermi liquids, not discussed here. For one-dimensional fermions see A.Luther and I. Peschel, Phys. Rev. B12, 3908, 1975 , S.Coleman Phys. Rev. D11, 2088, 1975. For the antiferromagnet see F.D.M.Haldane, Phys. Lett., Vol 93A, 464, 1983, and for the doped case see R.Shankar, Phys. Rev. Lett., Vol 63, 206, 1989. For the random Ising model see V.S. Dotsenko and V.S. Dotsenko, Adv. Phys., Vol 32, 129, 1983, B.N. Shalaev, Phys. Reports., {\bf 237}, 1994.  
\end{document}